%% file: ikchoo3.tex
\documentclass[letter]{ieice}
\usepackage[dvips]{graphicx}
\field{}
\title{Mutually Exclusive Procedures in Imperative Languages}
\authorlist{% fill arguments of \authorentry, otherwise error will be caused. 
\authorentry{Keehang KWON}{m}{labelA}
}
\affiliate[labelA]{The authors are with Computer Eng., DongA University. email:khkwon@dau.ac.kr}
\received{2003}{1}{1}
\revised{2003}{1}{1}
\finalreceived{2003}{1}{1}

\makeatletter
\long\def\@makemyfntext#1{$^{\rm *}\ $ #1}

\long\def\@myfootnotetext#1{\insert\footins{\footnotesize
    \interlinepenalty\interfootnotelinepenalty 
    \splittopskip\footnotesep
    \splitmaxdepth \dp\strutbox \floatingpenalty \@MM
    \hsize\columnwidth \@parboxrestore
   \edef\@currentlabel{\csname p@footnote\endcsname\@thefnmark}\@makemyfntext
    {\rule{\z@}{\footnotesep}\ignorespaces
      #1\strut}}}

\def\myfootnotetext{\@ifnextchar
     [{\@xfootnotenext}{\xdef\@thefnmark{\thempfn}\@myfootnotetext}}
\makeatother

\input macros

\newcommand{\muprolog}{{C$^{\uch}$}}
\newcommand{\uch}{uchoo}
\newcommand{\kch}{uchoo}

\begin{document}
\maketitle
\begin{summary}
%\begin{abstract}
To represent mutually exclusive procedures,  we propose
 a choice-conjunctive declaration statement of the form
$\uch(S,R)$ where $S, R$ are the procedure declaration statements within
 a module.
This statement has the
following semantics: request the machine to choose a successful one between $S$ and $R$. 
   This statement is useful for representing objects with mutually exclusive
procedures.

  We illustrate our idea
via  \muprolog, an extension of the core C  with a new 
 statement.
\end{summary}
\begin{keywords}
objects,  bounded choices, mutual exclusion.
\end{keywords}

\section{Introduction}\label{sec:intro}

 Despite  the attention, imperative languages \cite{Avi03,Jos12,Jos08} have
traditionally lacked mechanisms  for representing mutually exclusive tasks.
For example, an object like a coffee vending machine is in a superposition state of
mutually exclusive procedures, \ie, making a coffee or a tea and require further interactions to determine
their final task.

To represent objects with mutually exclusive procedures,  we propose to adopt a choice-conjunctive
operator in computability logic \cite{Jap03,Jap08}. To be specific,
we allow, within a module or class definition,
 a choice-conjunctive declaration statement of the form
$\uch(S_1,\ldots,S_n)$.
This statement has the
following semantics: request the machine to choose a successful one among $S_1,\ldots,S_n$.
  This statement  is useful for representing mutually exclusive tasks. 
Examples include 
function overloading or polymorphic procedures. For example,
the $switch$ field, declared as 

\[ \uch(switch == on,switch == off) \]

\noindent indicates that it
has two possible values, on and off,  and its final value will
be determined at run time by the machine.

%Another example is the $fibo$ procedure which typically has several implementations:
%a recursive version and a nonrecursive version.
Another example is the sorting procedure.

%\[ \uch(fib(n) = refib(n),  fib(n) = nonrefib(n)  \]

\[ \uch(qsort(L) = \ldots,  heapsort(L) = \ldots) \]

% kchoose(qsort(L) = ..., heapsort(L) = ...))
\noindent This system is in a superposition state of
several possible implementations and requires  the machine to determine
its final implementation.

It can be easily seen that our new statement has many applications
 in representing most interactive systems.
The following declaration represents an interactive object that requires
the machine to choose his major and  the amount of his tuition.

\begin{exmple}
module templeU \\
        \kch( \\
\>       major == english; tuition == \$2,000, \\
\>         major ==  medical; tuition == \$4,000,\\
\>           major == liberal; tuition == \$4,000);  \\
\end{exmple}

\noindent with the main program

\begin{exmple}
\>         read(major);  \\
 \>          print(tuition);
 \end{exmple}

\noindent 
In the above, the system 
 requests the user to type in a particular major. If the user  types in his major,
say, medical, then  the machine tries to select one 
among three majors, which leads to a success. After major == medical is selected, 
the machine sets his tuition to \$4,000 as well. The machine then prints the value of
the tuition.

This paper focuses on the minimum 
core of C. This is to present the idea as concisely as possible.
The remainder of this paper is structured as follows. We describe 
the core Java
 in Section 2. In Section \ref{sec:modules}, we
present an example of  \muprolog.
Section~\ref{sec:conc} concludes the paper.

%% </local definitions here>

\section{The Language}\label{sec:logic}

The language is a subset of the core C 
 with  procedure  definitions. It is described
by $G$- and $D$-formulas given by the syntax rules below:
\begin{exmple}
\>$G ::=$ \>   $true \sep A \sep x = v \sep  G;G    $ \\  
\>\>   \\
\>$D ::=$ \>  $ A = G\ \sep D; D \sep \all x\ D \sep \kch(D_1,\ldots,D_n)$ \\
\end{exmple}
\noindent
In the above, 
$A$  represents a head of an atomic procedure definition of the form $p(t_1,\ldots,t_n)$
or a field definition of the form $x == v$ where $x$ is a variable and $v$ is a simple value.
A $D$-formula  is called a  procedure definition. Note that a boolean condition is a legal
statement in our language.

In the transition system to be considered, $G$-formulas will function as the
main  statement, and a  $D$-formula enhanced with the
machine state (a set of variable-value bindings) will constitute  a program.
Thus, a program is a union of two disjoint sets, \ie, $\{ D \} \cup \theta$
where  $D$ is a $D$-formula and $\theta$ represents the machine state.
Note that $\theta$ is initially set to an empty set and will be updated dynamically during execution
via the assignment statements.

 We will  present an interpreter via a proof theory. % \cite{Khan87,MNPS91,HM94,MN12}
Note that this interpreter  alternates between 
 the execution phase 
and the backchaining phase.  
In  the execution phase (denoted by $ex(\Pscr,G,\Pscr')$) it tries to execute a main statement $G$ with respect to
a program $\Pscr$ and
produce a new program $\Pscr'$
by reducing $G$ 
to simpler forms until $G$ becomes an assignment statement or a procedure call. The rule (9) and (10) deal with this phase.
If $G$ becomes a procedure call or a boolean condition, the interpreter switches to the backchaining mode. This is encoded in the rule (8). 
In the backchaining mode (denoted by $bc(D,\Pscr,A,\Pscr')$), the interpreter tries 
to solve a procedure call  $A$ and produce a new  program $\Pscr'$
by first reducing a procedure definition $D$ in a program $\Pscr$ to simpler forms (via rule (3),(4),(5),(6))
 and then backchaining on the resulting 
definition (via rule (1) and (2)). 
 The notation $S$\ seqand\ $R$ denotes the sequential conjunctive execution of two tasks. To be precise, it denotes
the following: execute $S$ and execute
$R$ sequentially. It is considered a success if both executions succeed.
Similarly, the notation $S$\ parand\ $R$ denotes the parallel conjunctive execution of two tasks. To be precise, it denotes
the following: execute $S$ and execute
$R$ in parallel. It is considered a success if both executions succeed.

\begin{defn}\label{def:semantics}
Let $G$ be a main statement and let $\Pscr$ be a program.
Then the notion of   executing $\lb \Pscr,G\rb$ successfully and producing a new
program $\Pscr'$-- $ex(\Pscr,G,\Pscr')$ --
 is defined as follows:
\begin{numberedlist}

\item    $bc((p(t_1,\ldots,t_n) = G_1),\Pscr,p(t_1,\ldots,t_n),\Pscr_1)$ if
 $ex(\Pscr, G_1,\Pscr_1)$. \% A matching procedure for $p(t_1,\ldots,t_n)$ is found.

\item    $bc((x==v) = G_1),\Pscr,x==v,\Pscr_1)$ if
 $ex(\Pscr, G_1,\Pscr_1)$. \% a boolean conditional statement.

\item    $bc(\all x D,\Pscr,A,\Pscr_1)$ if   $bc([t/x]D,
\Pscr, A,\Pscr_1)$. \% argument passing

\item    $bc(D_1;D_2,\Pscr,A,\Pscr_1)$ if   $bc(D_1,
\Pscr, A,\Pscr_1)$. \% 

\item    $bc(D_1;D_2,\Pscr,A,\Pscr_1)$ if   $bc(D_2,
\Pscr, A,\Pscr_1)$. \% 

\item $bc(\kch(D_1,\ldots,D_n),\Pscr,A, \Pscr_1)$  if choose a successful one 
$bc(D_i,\Pscr,A, \Pscr_1)$ 

\item    $ex(\Pscr,A,\Pscr_1)$ if   $D \in \Pscr$ parand $bc(D,\Pscr, A,\Pscr_1)$. \% a procedure call or
a boolean condition.

%\item    $ex(\Pscr,x == E,\Pscr_1)$ if   $D \in \Pscr$ parand $bc(D,\Pscr, A,\Pscr_1)$. \% a procedure call

\item  $ex(\Pscr,true,\Pscr)$. \% True is always a success.

\item  $ex(\Pscr,x=E,\Pscr\uplus \{ \lb x,E' \rb \})$ if $eval(\Pscr,E,E')$.
\% the assignment statement. Here, 
$\uplus$ denotes a set union but $\lb x,V\rb$ in $\Pscr$ will be replaced by $\lb x,E' \rb$.

\item  $ex(\Pscr,G_1; G_2,\Pscr_2)$  if $ex(\Pscr,G_1,\Pscr_1)$  seqand 
  $ex(\Pscr_1,G_2,$ \\ $\Pscr_2)$. \% sequential composition

\end{numberedlist}
\end{defn}

\noindent
If $ex(\Pscr,G,\Pscr_1)$ has no derivation, then the machine returns  the failure.

\section{Examples }\label{sec:modules}

As an  example, consider a simple smartphone which performs only two  mutually exclusive tasks.  
The types of smartphone tasks  are 1) play music with the speaker on, and, 2) sleep with the
speaker off.
An example of this object is provided by the
following code where the program $\Pscr$ is of the form:

\begin{exmple}
module smartphone \\
 \kch(speaker == on,
      speaker == off);  \\
playmusic(x) = speaker == on; play music x hours; \\
sleep(y) =  speaker == off; sleep y hours \\
\end{exmple}

\noindent and the goal $G$ is of the form:

\begin{exmple}
        while true \\
        playmusic(10); \\
        sleep(14); \\
        endwhile;
\end{exmple}
\noindent In the above,  the machine plays the music for ten hours by turning
on the speaker. After ten hours of playing, the machine sleeps for fourteen hours
by turning off the speaker.
 Then the execution will repeat it again.

\section{Conclusion}\label{sec:conc}

In this paper, we extend  the core C with the addition of
conjunctive statements within a class definition. This extension allows statements of
the form  $\uch(D_1,\ldots,D_n)$  where each $D_i$ is a definition statement.
This statement makes it possible for the core C
to model decision steps from the machine. 

\section{Acknowledgements}

This work  was supported by Dong-A University Research Fund.

\bibliographystyle{ieicetr}

%\profile*{}{}% without picture of author's face

\end{document}

%% file: macros.tex
\newenvironment{numberedlist}
{\begin{list}{\makebox[20pt]{\hss(\arabic{itemno})\enspace}}
             {\usecounter{itemno}\labelwidth 20pt}}{\end{list}}

\newcounter{itemno}

\newcounter{itemno1}

\newcounter{itemno2}

\newcounter{exno}

\newcounter{defno}

\newenvironment{defn}{\refstepcounter{defno}\medskip \noindent {\bf
Definition \thedefno.\ }}{\medskip}

\newcommand{\sep}{\;\vert\;}

\newcommand{\oprove}{\vdash\kern-.6em\lower.7ex\hbox{$\scriptstyle O$}\,}

\newcommand{\Pscr}{{\cal P}}

\newcommand{\pderivation}{{\cal P}\kern -.1em\hbox{\rm -derivation}}
\newcommand{\pderivationl}{{\cal P}\kern -.1em\hbox{\em -derivation}}
\newcommand{\pderivable}{{\cal P}\kern -.1em\hbox{\rm -derivable}}
\newcommand{\pderivablel}{{\cal P}\kern -.1em\hbox{\em -derivable}}
\newcommand{\pderivations}{{\cal P}\kern -.1em\hbox{\rm -derivations}}
\newcommand{\pderivability}{{\cal P}\kern -.1em\hbox{\rm -derivability}}

\newcommand{\all}{\forall}

\newcommand{\ie}{{\em i.e.}}

\newsavebox{\lpartfig}
\newsavebox{\rpartfig}

\newenvironment{exmple}{
 \begingroup \begin{tabbing} \hspace{2em}\= \hspace{3em}\= \hspace{3em}\=
\hspace{3em}\= \hspace{3em}\= \hspace{3em}\= \kill}{
 \end{tabbing}\endgroup}

\newcommand{\lb}{\langle}
\newcommand{\rb}{\rangle}

     {\\* \hspace*{\fill} \end{trivlist}}

%% file: ikchoo3.bbl
\begin{thebibliography}{1}


\bibitem{Jap03}
G.~Japaridze, ``Introduction to computability logic'', Annals  of Pure and
 Applied  Logic, vol.123, pp.1--99, 2003.

\bibitem{Jap08}
G.~Japaridze,   ``Sequential operators in computability logic'',
 Information and Computation, vol.206, No.12, pp.1443-1475, 2008.


\bibitem{Lyn96}
Lynn Andrea Stein, ``Interactive programming: revolutionizing introductory computer science," ACM Comput. Surv. 28, 4es, Article No. 103, December 1996.

\bibitem{Rol10}
Roly Perera, ``First-order interactive programming," 12th international conference on Practical Aspects of Declarative Languages (PADL'10),
Springer-Verlag, Berlin, Heidelberg, pp. 186-200, Jan. 2010.

\bibitem{Avi03}
Avinash C. Kak, Programming with Objects: A Comparative Presentation of Object-Oriented Programming with C++ and Java,
John Wiley \& Sons, Inc., New York, NY, USA, 2003.

\bibitem{Jos12}
Joseph Albahari, and Ben Albanhari, C\# 5.0 Pocket Reference, O'Reilly Media, 224 pages, May 2012.

\bibitem{Jos08}
Joshua Bloch, Effective Java, Second Edition, Addison-Wesley, 346 pages, May 2008.

\end{thebibliography}
